\documentclass[preprintnumbers,aps]{revtex4}
\usepackage{graphicx}
\usepackage{amssymb,amsmath}
\usepackage{dcolumn}
\usepackage{bm}
\usepackage{color}
\usepackage{hyperref}
\begin{document}
\title {Principles of Control for Decoherence-Free Subsystems}
\author{P. Cappellaro\footnote{These authors contributed equally to this work.}, J. S. Hodges$^*$, T. F. Havel and D. G. Cory\footnote{Author to whom correspondence should be addressed. Electronic mail: dcory@mit.edu}}
\affiliation{Massachusetts Institute of Technology, Department of Nuclear Science and
Engineering, Cambridge, MA 02139, USA\bigskip}
\newcommand{\half}{\frac{1}{2}}
\newcommand{\mket}[1]{\vert{)1}\rangle}
\newcommand{\mbra}[1]{\langle{)1}\vert}
\newcommand{\ket}[1]{\vert{#1}\rangle}
\newcommand{\bra}[1]{\langle{#1}\vert}
\newcommand{\ham}{\mathit{H}}
\newcommand{\Tr}[1]{\textrm{Tr}\left[{#1}\right]}
\graphicspath{{../CorrelationTim/}}
\begin{abstract}
Decoherence-Free Subsystems (DFS) are a powerful means of protecting quantum information against noise with known symmetry properties. Although Hamiltonians theoretically exist that can implement a universal set of logic gates on DFS encoded qubits without ever leaving the protected subsystem, the natural Hamiltonians that are available in specific implementations do not necessarily have this property.  Here we describe some of the principles that can be used in such cases to operate on encoded qubits without losing the protection offered by the DFS. In particular, we show how dynamical decoupling can be used to control decoherence during the unavoidable excursions outside of the DFS. By means of cumulant expansions, we show how the fidelity of quantum gates implemented by this method on a simple two-physical-qubit DFS depends on the correlation time of the noise responsible for decoherence. We further show by means of numerical simulations how our previously introduced ``strongly modulating pulses'' for NMR quantum information processing can permit high-fidelity operations on multiple DFS encoded qubits in practice, provided that the rate at which the system can be modulated is fast compared to the correlation time of the noise. The principles thereby illustrated are expected to be broadly applicable to many implementations of quantum information processors based on DFS encoded qubits.
\end{abstract}
\maketitle

\section{Introduction}
The interaction of a quantum system with its environment leads to the loss of quantum phase information and interference, which is widely known as decoherence \cite{Zurek:91, GiuliniEtAl:96, BreuerPetruc:02}. This conversion of quantum information into classical information is a major obstacle in any application where pure quantum phenomena are sought, particularly quantum information processing (QIP).
One promising strategy for countering decoherence is to encode the information in subsystems that, because of the symmetry of the interactions between the qubits and their environment, are invariant under the action of the noise generators \cite{DFSZanardi,DFSGuo}. 

Universal computation within these decoherence-free subsystems (DFS) has been shown to be possible from a theoretical point of view \cite{viola,ViolaDFS,bacon}. In particular, universal fault-tolerant computation within a DFS is possible if the exchange interaction between qubits can be switched off and on at will \cite{DFSuniversal2,DFSuniversal3,XYintNature}. The issue that we address here arises in systems where this Hamiltonian does not occur naturally.  

The total Hamiltonian  is conveniently divided into a time-independent part, $\mathit{H}_{int}$, and a part that depends on a set of experimentally controllable, time-dependent parameters, $\mathit{H}_{ext}$({$\alpha_i$(t)}).  There are many useful DFS encodings for which the generators of logical qubit rotations are not contained in the set of available total Hamiltonians: 
$\mathit{H}_{tot}= \mathit{H}_{int}+\mathit{H}_{ext}$.  As long as the system is universal, any desired propagator on the logical subsystem can be composed from the evolution under a series of time dependent external Hamiltonians; the instantaneous total Hamiltonian need not preserve the protected subsystem.   In such cases, the extent to which universal fault-tolerant computation is possible depends on the details of the control fields, as well as the \emph{spectral density} of the noise \cite{Uchiyama}. This paper will discuss these issues in the context of a specific QIP implementation based on liquid-state NMR.

 In Section II it is shown that the radio-frequency (RF) control fields used in NMR necessarily cause the encoded information to ``leak'' from the protected subsystem into other parts of the total system Hilbert space, where it is subject to decoherence.
This is illustrated by numerical simulation of a simple example -- the encoding of one logical qubit in a two-spin  decoherence-free subspace for collective dephasing \cite{DFSevan} -- which is referred to throughout the paper. It is further shown that in the case of two DFS qubits, each with the same noise model, the leakage rate is generally nonzero even  in the absence of the control Hamiltonian.

In Section III, we briefly review the application of stochastic Liouville theory \cite{GhoseAvgLiouvilleTheory,ChengSilbey:04} to understand the effective decoherence in the presence of an external Hamiltonian that modulates the spin dynamics, as is the case for strongly-modulating pulses (SMP) \cite{softpulses,PBEFFHMC:03} or optimal control theory \cite{NavinGRAPE}, dynamical decoupling \cite{DynDecViolaKnill} or bang-bang control \cite{BBLloyd} .
We then use these results in Section IV to quantitatively understand the Carr-Purcell \cite{CP} (CP) sequence.  CP is perhaps the original dynamical decoupling sequence and the archtype for observing the influence of the noise correlation time on the effective decoherence rate.

We will show that the CP sequence can be effective at suppressing both leakage and decoherence provided that one can modulate
 the system on a time scale shorter than the correlation time of the noise.
Under the assumption of instantaneous and perfectly selective single spin $\pi$-pulses, the exact dependence of the overall gate fidelity on the pulse rate and the correlation time of the noise is derived for a single DFS qubit.
 Finally in Section V, we apply these ideas to a physical system with realizable models of control fields.  This includes limitations on the available RF power and the lack of frequency selectivity among the physical spins. As expected, pulses of finite duration degrade the gate fidelities of these operations.
Analytical solutions are generally not feasible, and hence the amount of degradation that can be expected for a range of experimentally realistic parameters is evaluated by means of numerical simulations for simple quantum gates operating upon one or two encoded qubits.
{Lastly, we show the criteria of noise correlation times that are compatible with high fidelity control.}

\section{Leakage from a two-logical qubit DFS}
In  physical implementations of QIP, qubits are not embodied in well-isolated two-state systems: rather, they are embedded in larger Hilbert spaces containing additional states  that are intended to not participate in the computation.
In addition, when physical two-state systems are combined into logical qubits for noise protection and/or correction, additional redundant degrees of freedom are introduced.
 Leakage to these ``external'' degrees of freedom can destroy the coherent dynamics of the qubit \cite{LeakageLloydPRA,LeakageLidarPRL}. 

Such leakage can be introduced by the control fields applied to implement specific logic gates, and even in absence of external fields, by the internal Hamiltonian itself.   Many modulation methods have been engineered to refocus the terms in the internal Hamiltonian responsible for leakage \cite{LeakageLidarPRA}. We explore the fidelity that can be reasonable  expected based on the details of the modulation scheme and the spectral density of the noise.

In this section we use a simple well-known DFS to motivate our discussion. The DFS   encodes one logical qubit in two physical spin-$\half$ particles  and protects against collective dephasing caused by fully correlated uni-axial noise.  In NMR, for example, fluctuations of the quantizing magnetic field $B_z$ at a local molecule appear fully correlated, yet lead to dephasing when averaged over the spin ensemble \cite{DFSevan}.
These fluctuations are described by a  Hamiltonian of the form $\mathit{H}_{SE}(t) = \gamma B_z(t) \, \mathit{Z}$, where $\mathit{Z} =  \half\sum_i \sigma_z^i$ is the total angular momentum of the spins along the $z$-axis and $\gamma$ is their gyromagnetic ratio. 
The DFS is based on the encoding $\ket{0}_L=\ket{01},\ \ket{1}_L=\ket{10}$.
A basis for the space of operators on the encoded qubit, in turn, is given by the four logical Pauli operators:
\begin{align}
	\label{EncodedOperations}
		\sigma_z^L & ~\Leftrightarrow~ \tfrac12 \big( \sigma_z^1-\sigma_z^2 \big) & \sigma_x^L & ~\Leftrightarrow~ \tfrac12 \big( \sigma_x^1\sigma_x^2+\sigma_y^1\sigma_y^2 \big) \\ \notag
		\openone^{L} & ~\Leftrightarrow~ \tfrac12 \big( \openone^{1,2}-\sigma_z^1\sigma_z^2 \big) & \sigma_{y}^L & ~\Leftrightarrow~ \tfrac12 \big( \sigma_x^1\sigma_y^2-\sigma_y^1\sigma_x^2 \big)
\end{align}

This two spin-$\half$ particle Hilbert space ($\mathbb{C}^4 = \mathbb{C}^2 \otimes \mathbb{C}^2$) can be described as a direct-sum of the total angular momentum subspaces, $Z_{0} \oplus Z_{+1} \oplus Z_{-1}$, where $l$ is the total angular momentum projected along the quantization axis.  The logical basis states  $\ket{0}_L$ and $\ket{1}_L$ reside exclusively in $Z_{0}$, where $Z_{0} \equiv \mathbb{C}^2$.  When we discuss leakage, we imply that the instantaneous state in $\mathbb{C}^4$ has elements in $Z_{\pm1}$.  In this case, the information within the state of the system cannot be described completely by the four operators above (Eq. \ref{EncodedOperations}).  Since the total angular momentum with $l = 0$ is a constant of the motion under the system-enviroment Hamiltonian, a state not completely represented by a linear combinations of (Eq. \ref{EncodedOperations}) is thus affected by decoherence.  We will expore this DFS as implemented in liquid state NMR for both one and two logical qubits. 

The internal Hamiltonian (in the rotating frame) for two spins in liquid-state NMR already is exclusively in $Z_0$ and thus can be expressed by the operators in Eq. \ref{EncodedOperations}; it does not cause mixing of the subspaces $Z_l$,
\begin{eqnarray}
\label{Ham2spins}
	\mathit{H}_{int} ~=~ \tfrac{\Delta\omega_{12}}{2}(\sigma_z^1-\sigma_z^2) + 
	\tfrac{\pi}{2}\, J_{12}\, \vec{\sigma} ^1\cdot\vec{\sigma}^2 , 
\end{eqnarray}
where $\Delta\omega_{12}$ is the difference in chemical shift of the two spins and $J_{12}$  the scalar coupling constant.
The former coefficient scales the logical $\sigma_z^L$ operator, while the latter scales the $\sigma_x^L$ operator.
Thus evolution under the internal Hamiltonian alone generates a continuous rotation about an axis in the logical $xz$-plane making an angle of $\arctan( \pi J_{12}/\Delta\omega_{12})$ with the logical $x$-axis.
As illustrated below, more general gates can be obtained via the interplay of the internal Hamiltonian and an external time-dependent RF (radiofrequency) field, with Hamiltonian
\begin{equation}
\label{RF}
\mathit{H}_{ext} ~=~ \omega_{\textsf{rf}}(t)\, e^{-i Z\phi(t)} X e^{\,i Z\phi(t)} \quad \left( X ~\equiv~ {\textstyle\frac12\,\sum}_i\, \sigma_x^i \right) ~,
\end{equation}
where $\phi(t)$ is a time-dependent phase and $\omega_{\textsf{rf}}(t)$ is a time-dependent amplitude.  The phase and amplitude are independently controllable.
Note that $H_{ext}$ cannot be expressed as a linear combination of the logical Pauli operators.  In the presence of RF fields the evolution of a state inside the DFS under  the total Hamiltonian necessarily causes the information to ``leak'' outside of the DFS, where it is no longer immune to collective dephasing.  This combination of $\mathit{H}_{int}$ and $\mathit{H}_{ext}(t)$ can generate any unitary in the $\mathbb{C}^4$ Hilbert space, guaranteeing universality.  In the absence of decoherence and assuming ideal controls, we can reach a unit fidelity for any desired gate \cite{khanejaOCT,RabitzOTC}.  Our interest is assessing the fidelity of control for a finite decoherence, for a finite bandwidth of our control parameters, and for Hamiltonians not respecting the symmetry of the logical subspace.

\begin{figure}[hbt]
		\includegraphics[scale=0.35]{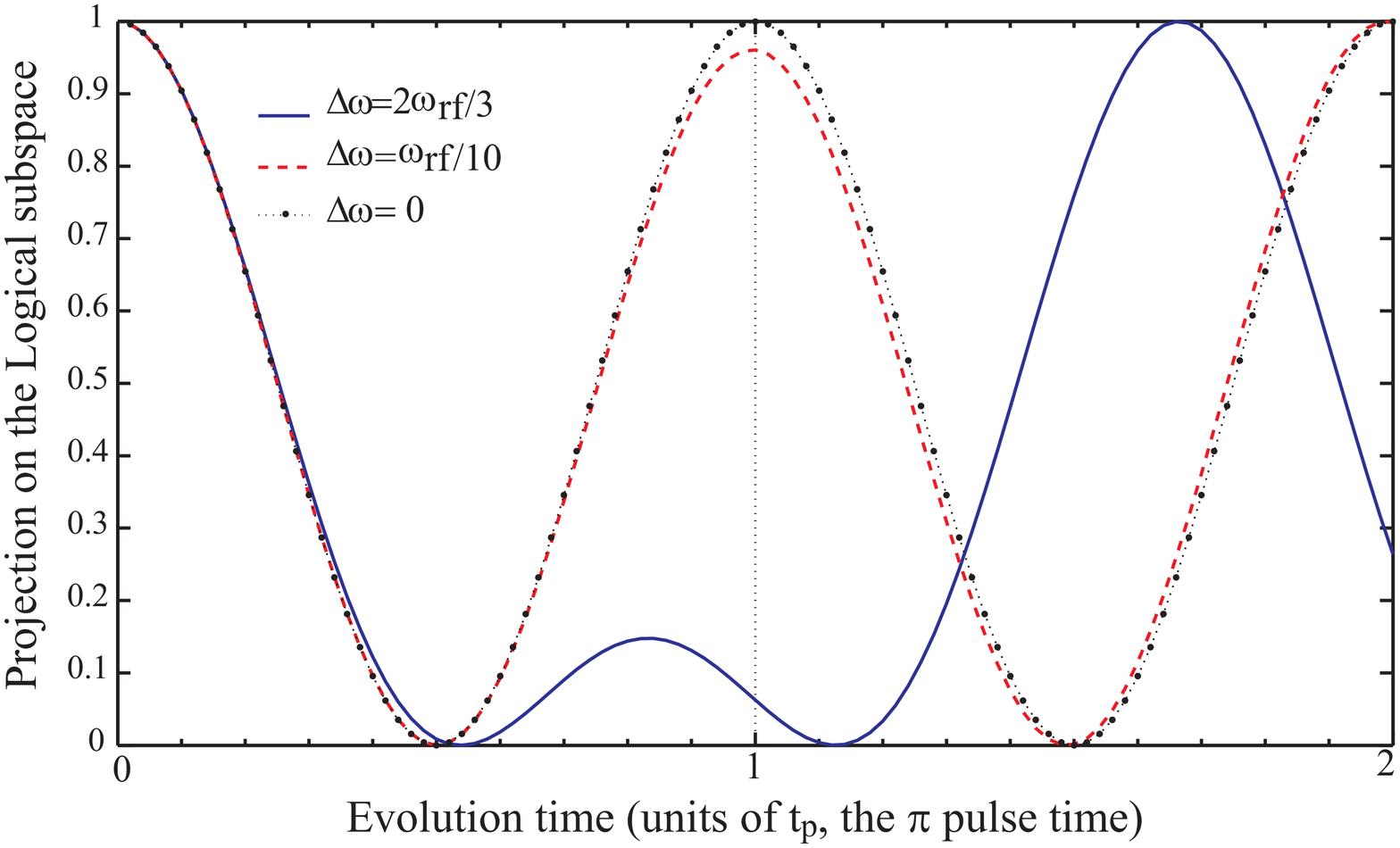}
\caption{
Shown above is the projection onto the logical subspace of a state initially inside the DFS, during application of an RF pulse for various ratios of $\frac{\Delta\omega}{\omega_{\textsf{rf}}}$. Defining the projection operator onto the logical subspace as  $P_L$ , we plot $p(t)=Tr[(P_L\rho(t))^2]/Tr(\rho(t)^2)$, for $t =0\rightarrow2t_p$, where $\rho(t)=e^{-i\omega_{\textsf{rf}}t(\sigma_x^1+\sigma_x^2)}\sigma_z^Le^{i\omega_{\textsf{rf}}t(\sigma_x^1+\sigma_x^2)}$ and $\omega_{\textsf{rf}}t_p=\pi$. The logical state completely returns to the subspace  after application of a $\pi$-pulse  to both spins only when the  spins have identical resonance frequencies ($\Delta\omega=0$). If the ratio $\frac{\Delta\omega}{\omega_{\textsf{rf}}}$ is non-zero, as required for universality, the return to the logical subspace is imperfect (in particular, it is in general possible to go back to a state very close to the initial state in a time $t>t_p$, but it is much more difficult to implement a $\pi$ rotation). A logical $\pi$-pulse using a single period of RF modulation is not possible, a more complex RF modulation, like composite pulses \cite{LevittComposite}, strongly-modulating pulses \cite{softpulses,PBEFFHMC:03} or optimal control theory \cite{NavinGRAPE}, is required. In the above model, $\frac{\omega_{\textsf{rf}}}{J}=500$; the initial state of the system is $\sigma_z^L$}
\label{fig:pipulseleakage}
\end{figure}

  In the case of ideal control fields, an instantaneous $\pi$-pulse ($t_p \rightarrow 0$)  corresponds to a logical operation \cite{DFSevan}, since $P_x(\pi)=e^{-i\pi/2(\sigma_x^1+\sigma_x^2)}=-e^{-i\pi/2(\sigma_x^1\sigma_x^2)}$, which is equivalent to a $\pi$ pulse around $\sigma_x^L$. Figure \ref{fig:pipulseleakage} motivates the extent to which universality within the subsystem can be obtained in the finite $t_p$ regime.  In this figure, we plot the purity of the projection of $\rho(t)=e^{-i\omega_{\textsf{rf}}t(\sigma_x^1+\sigma_x^2)}\sigma_z^Le^{i\omega_{\textsf{rf}}t(\sigma_x^1+\sigma_x^2)}$ on the logical subspace.  In the limit of very high RF power ($\frac{\Delta \omega}{\omega_\textsf{rf}} \rightarrow 0$), the system  undergoes a $\pi$-pulse in a time $t_p = \frac{\pi}{\omega_\textsf{rf}}$ and returns completely to the subspace after this time. It remains outside the subspace only for the duration of the pulse.  For $\omega_\textsf{rf}$ which are physically relevant ($\omega_1 < 2\pi 100$ kHz and $0<\Delta\omega<2\pi 20$ kHz),
 a single RF pulse does not result in a logical $\pi$-rotation due to off-resonance effects.  Experimentally we are limited to finite $t_p$ and even our simple two logical qubit model system is sufficient to introduce several key challenges in implementing coherent control over logical qubits: (i) decoherence due to leakage outside the subspace during RF modulation periods, (ii) decoherence due to leakage outside the subspace after RF modulation, and (iii) loss of fidelity due to cumulative leakage with respect to the spectral density of the noise.

\begin{figure}[htb]
		\includegraphics[scale=0.4]{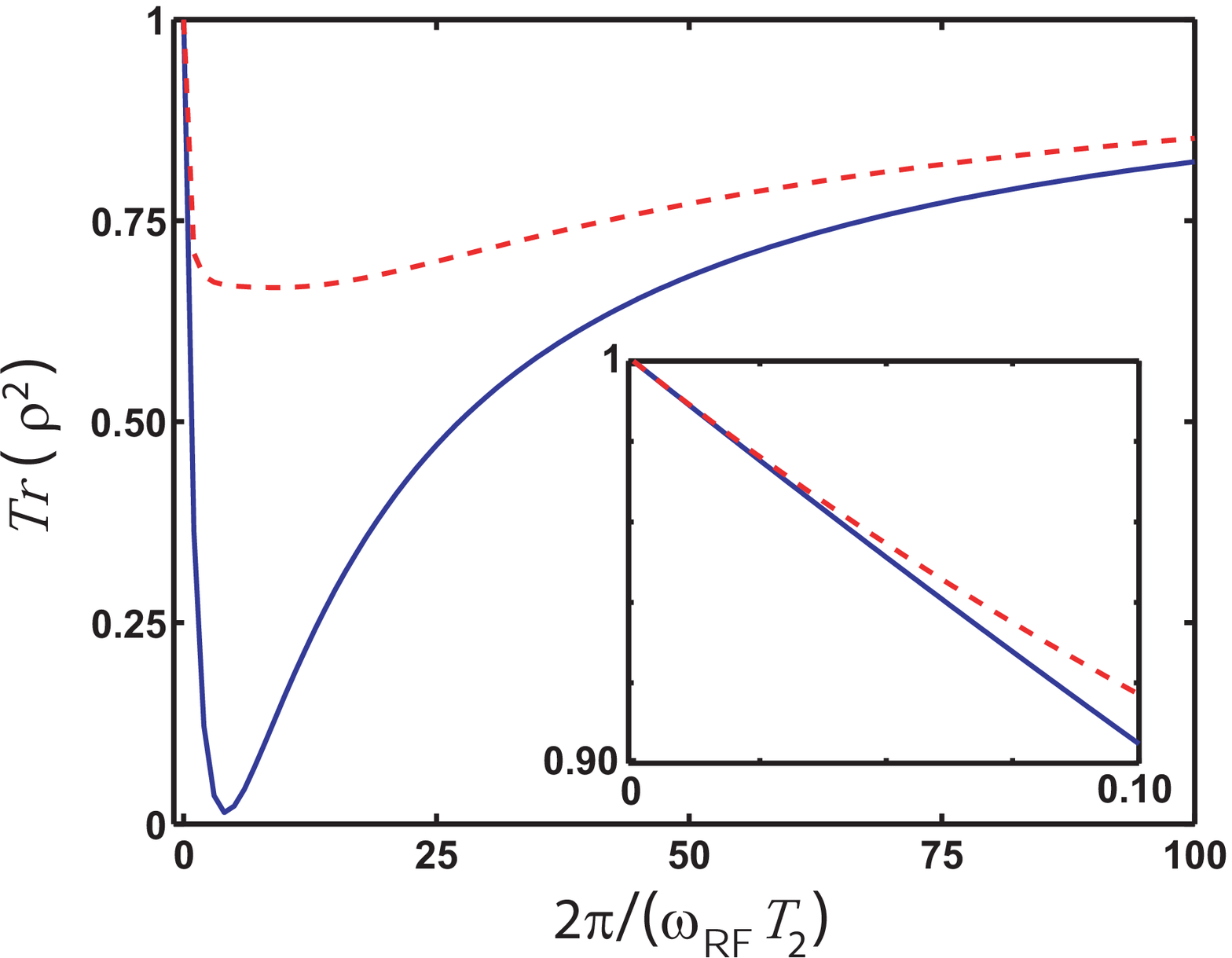}
		\includegraphics[scale=0.4]{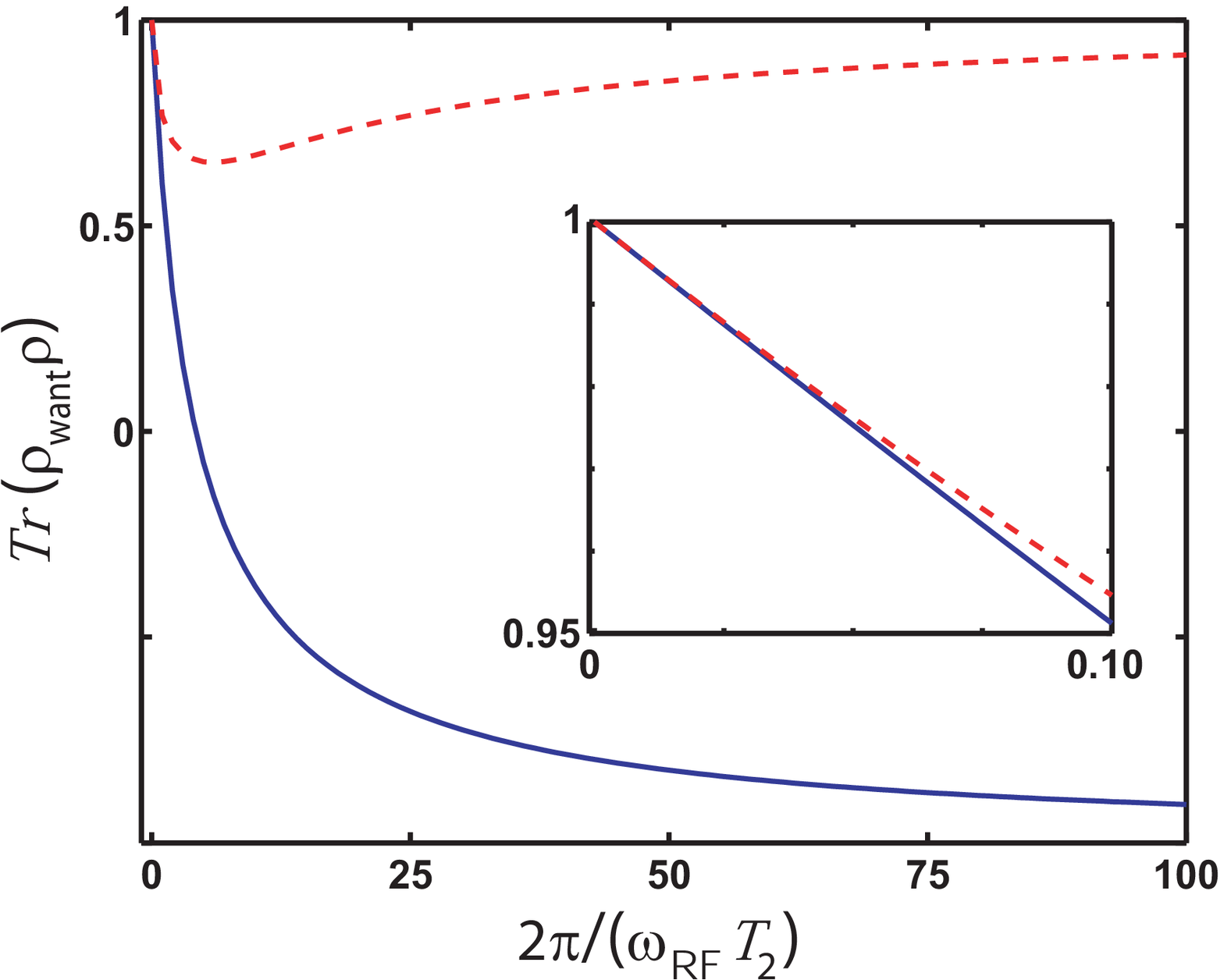}
\caption{%
Plots illustrating the loss of fidelity due to totally correlated decoherence during the application of a $\pi$-pulse about the $x$-axis to the two spins of the DFS (see text).
The dashed curves (red in the on-line version) are for the initial states $\rho_0 =\openone^L$ or $\rho_0 = \sigma_x^L$, while the lower curves (blue in the on-line version) are for $\rho_0 = \sigma_y^L$ or $\rho_0 = \sigma_z^L$.
The left-hand plot shows the trace of $\rho^2$ following the $\pi$-pulse as a function of the inverse product of the RF power $\omega_{\textsf{rf}}$ and the relaxation time $T_2$.
The right-hand plot shows the correlation with the ideal final state, i.e.~the trace of $\rho_{\textsf{want}} \rho$, following the $\pi$-pulse as a function of this same parameter.
} \label{fig:pipulseplots}
\end{figure}

Figure \ref{fig:pipulseplots} shows an illustrative example of the integrated effects of a $\pi$-pulse applied to the two spins in such a DFS on the purity ($\Tr{\rho^2}$) and correlation with the ideal final state ($\Tr{\rho_{\textsf{want}} \rho}$) as a function of the ratio of the relaxation rate $1/T_2$ to the RF power $\omega_{\textsf{rf}}$.
The initial states were chosen from the four logical Pauli operators, and we made the approximation that the internal Hamiltonian is zero during the application of these $\pi$-pulses.
As would be expected, the desired result (negating the state in the case of $\rho_0 = \sigma_y^L, \sigma_z^L$, or preserving it for $\rho_0 = \openone^L, \sigma_x^L$) is rapidly degraded by the totally correlated decoherence during the $\pi$-pulse, unless the Rabi frequency is considerably faster than the relaxation rate.
The increase in both the coherence and the correlation when the relaxation becomes fast compared to the rotation rate is due to a sort of ``quantum Zeno'' effect, so that the RF field itself is unable to rotate the state out of the DFS.
In a complete analysis of the 2-spin case the effects shown in Fig. \eqref{fig:pipulseleakage}  must be combined to those in Fig. \eqref{fig:pipulseplots}.

Manipulating more than one logical qubit introduces further complexities to the control versus leakage problem. 
For the DFS considered, the extension to 2 logical qubits encoded into 4 physical qubits leads to the following basis states:
\begin{equation}
	\label{encoding2}
	\begin{array}{ll}
		|00\rangle_L \Leftrightarrow |01 01\rangle, \ \ \ &\ \ \  |10\rangle_L \Leftrightarrow  |10 01\rangle, \\
		|01\rangle_L \Leftrightarrow |01 10\rangle, \ \ \ &\ \ \  |11\rangle_L \Leftrightarrow  |10 10\rangle
	\end{array}
\end{equation}

We define leakage any evolution that will cause the state to be not fully described by a linear combination of this four basis vectors.
Since we focus on the challenges unique to controlling multiple logical qubits, we assume the internal Hamiltonian to be given by the Hamiltonians of each logical pair (as in eq. \eqref{Ham2spins}) and a coupling between two spins pertaining to two distinct logical qubits:
\begin{equation}
	\label{Ham4_23}
	\ham=\ham_{L1}+\ham_{L2}+\tfrac12 \pi J_{23}\, \vec\sigma_2 \cdot \vec\sigma_3
\end{equation}
The interaction term $\ham^I_{23}=\vec\sigma_2\cdot\vec\sigma_3$ couple the system initially in the subspace defined by the state \eqref{encoding2} to the subspace defined by the states 
\begin{equation} \label{additionalzq}
	\ket{0011}\ \ \ \ \text{and}\ \ \ \ \ket{1100},
\end{equation}
for example $e^{-i\pi/4\ham^I_{23}}\ket{0101}=\frac{1-i}{\sqrt{2}}\ket{1100}$. 
If the noise is collective only over each pair of spins that encodes a logical qubit \cite{viola2encdyndec}, the states \eqref{additionalzq} are not protected against it and will decohere. The internal Hamiltonian will be therefore responsible for leakage and the ultimate decay of the system. 

Notice that we would in general expect the noise to be collective over all the physical qubits, and not pairwise collective. In the case of NMR, this corresponds to a fluctuating external magnetic field, which is fully correlated. However, the differences in energies between qubits could be strong enough to effectively add a non-collective component to the noise. In particular, we can consider in NMR the case in which each pair is formed by spins of a different chemical species. In this case, the difference in gyromagnetic ratio makes the strength of the noise acting on each pair unequal, so that the noise is no longer collective. On the other hand, when the Zeeman energy separation is considerable, the coupling between spins can be very well approximated by the diagonal part of $\ham^I_{23}$, i.e. $\sigma_z^2\sigma_z^3$, which does not cause leakage.

When the noise generator is fully collective (as for homonuclear systems in NMR), the internal Hamiltonian still causes leakage, via the coupling to the states \eqref{additionalzq}. Since these states belong to the zero eigenvalue subspace of the noise generator, they do not decohere. Information could still be lost at the measurement stage, since the states \eqref{additionalzq} are not faithfully decoded to a physical states. A unitary operation is enough to correct for this type of leakage, and since decoherence is not an issue here, there are no concerns regarding the time scale over which the correction should be applied; however, amending for this unwanted evolution would in general mean the introduction of an external control, that, as seen, is a source of leakage leading to decoherence.
 
For logical encodings other than the DFS considered, the natural Hamiltonian may drive the state out of the protected subspace even for single logical qubits; for example, the \textit{noiseless subsystem} considered in Ref.~\cite{noiseless} will evolve out of the protected subspace whenever the chemical shifts or scalar couplings among its three constituent spins are not all equal.

If we wish to do something more complicated than merely freeze the evolution of the system, e.g.~to rotate the DFS qubits while simultaneously refocusing all the interqubit couplings, the complexity of the modulation sequence increases and the various causes of leakage will combine.
In attempting to demonstrate a universal set of logic gates on a pair of two-spin DFS qubits by liquid-state NMR, leakage turned out to be an unavoidable problem for all practical intents and purposes.
Fortunately, it turns out that in many practical situations other means of inhibiting decoherence are also available, and can allow one to leave the protected subspace if need be in order to simplify the implementation of logic gates on encoded qubits.
Dynamical decoupling is a particularly promising class of techniques for these purposes, which are applicable whenever the correlation time of the noise is long compared to the rate at which the system can be coherently modulated.
The next section will analyze the principles involved in this approach, and show how they may be applied to some simple but realistic examples. 

\section{Stochastic Liouville Theory and Cumulant Averages}
The earliest example of a pulse sequence that could correct for random field fluctuations with long correlation times was given by Carr and Purcell \cite{CarrPurcell:54,MeiboomGill:58,Freeman:98}. 
Today this would be regarded as dynamical decoupling or ``bang-bang'' control \cite{violaBB} and it has been applied beyond magnetic resonance, for example to the control of decoherence in spin-boson models \cite{UchiAihar:03}.
In this section we outline a formalism, based on the well-known stochastic Liouville formalism \cite{Levanon, HavelEtAl:01, ChengSilbey:04} and cumulant expansion, which allows us to analyze the effects of dynamical decoupling on decoherence. In the following section we apply this to the above  two-spin DFS.

Stochastic Liouville theory is based on a semiclassical model of decoherence, in which the Hamiltonian at any instant in time consists of a deterministic and a stochastic part.
In the simplest case of NMR $T_2$ relaxation, this typically takes the form
\begin{equation}
	\label{totHam}
	\mathit{H}_{tot}(t) ~=~ \mathit{H}_{det}(t) + \mathit{H}_{st}(t) ~=~ \mathit{H}_{int} + \mathit{H}_{\textsf{rf}}(t) + {\sum}_k \omega_k(t) \mathit{Z}_k ~,
\end{equation}
where $H_{int}$ is the static internal Hamiltonian, $H_{\textsf{rf}}(t)$ is the RF Hamiltonian, the $\omega_k(t)$ describe the phase shifts due to stochastic, time-dependent fluctuating fields and $\mathit{Z}_k$ are the generators of each of these noise sources, i.e.~operators which describe how these classical fields are coupled to the quantum system.
In the two-spin DFS example considered previously, there is only one noise generator $\mathit{Z} = (\sigma_z^1 + \sigma_z^2)/2$ with $\omega(t) = \gamma B(t)$, which describes collective fluctuations parallel to the applied static magnetic field.

We now introduce a superoperator $\mathcal L(t)$ defined on Liouville (operator) space via
\begin{equation}
 \label{supOp}
   \mathcal L(t) ~=~ \mathit{H}_{tot}^{\,*}(t)\otimes \openone - \openone \otimes \mathit{H}_{tot}(t) ~=~ \mathcal L_{det}(t) \,+\, {\sum}_k \omega_k(t) \mathcal Z_k
\end{equation}
where $\mathcal Z_k = Z_k^* \otimes \openone - \openone \otimes Z_k$.
This superoperator is the generator of motion for density operator $\hat\rho$, meaning
\begin{equation}
	\label{evolution1}
	{\rho}(t) ~=~ \mathcal U\, \hat{\rho}(0) ~=~ \mathcal T \exp\!\bigg( -i \int_0^t dt'\, \mathcal L(t') \bigg)\, \hat{\rho}(0)
\end{equation}
where $\mathcal T$ is the usual time ordering operator. 
Since what is actually observed in an experiment is the statistical average over the microscopic trajectories of the system $\langle \hat{\rho}(t) \rangle$, we have to take the ensemble average superpropagator to obtain $\langle \hat{\rho}(t) \rangle=\big\langle \mathcal U \big\rangle \hat{\rho}(0)$.
The problem of calculating the average of the exponential of a stochastic operator has been solved by Kubo \cite{kubo} using the cumulant expansion.
In terms of the so-called ``cumulant averages'' $\langle\,\cdots\rangle_c$ (see Appendix A), the superpropagator is given by:
\begin{equation}
	\label{cumulant4}
	\big\langle \mathcal U \big\rangle ~=~ \exp\! \left(-\,i \int_0^t dt'\, \langle \mathcal L(t')\rangle_c- \tfrac12\, \mathcal T\! \int_0^t dt_1 \int_0^t dt_2\, \langle \mathcal L(t_1) \mathcal L(t_2) \rangle_c
	    +\cdots \right)
\end{equation}
Providing $\|\int_0^t dt'\mathcal L(t')\| \ll 1$ for all $t > 0$ we can safely neglect high order terms in the exponential's argument.

Similar expressions are obtained in the formalism of average Hamiltonian theory (AHT) \cite{AHTHaberlen} for the coherent (instead of stochastic) averaging of the system evolution under control Hamiltonians cyclic and periodic in time. Here we can obtain simplifications analogous to those encountered in AHT if we analyze the evolution in the interaction frame (called ``toggling frame'' in NMR) defined by the RF propagator $U_{\textsf{rf}}(t)$ \cite{HabCohAve}.
In this frame the noise operators acquire a further time-dependency (coherently imposed by the cyclic excitation) in addition to the stochastic time dependency of their coefficients $\omega_k(t)$.
The total Hamiltonian in the toggling frame is
\begin{equation}
	\label{totHamTog}
	\tilde{\mathit{H}}_{tot}(t) ~=~ \tilde{\mathit{H}}_{det}(t) + {\sum}_k \omega_k(t)\tilde{\mathit{Z}}_k(t),
\end{equation}
where the toggling frame equivalent $\tilde{O}$ of any given operator $O$ is defined by $\tilde{O}(t) ~=~ U_{\textsf{rf}}^\dag(t)\, O\, U_{\textsf{rf}}(t)$, with :
\begin{equation}
	\label{toggling}
		U_{\textsf{rf}}(t) ~\equiv~ \mathcal T \exp\!\bigg(-i\int_0^t dt'\, \mathit{H}_{\textsf{rf}}(t') \bigg), 
\end{equation}
and  $U_{\textsf{rf}}(t_c)=\openone$ for cyclic controls, so that the toggling frame and laboratory frame coincide at the end of each cycle.
 
This time-dependent change of basis in Liouville space induces a change of basis in the space of superoperators acting on Liouville space, as a result of which the noise super-generators $\mathcal Z_k$ also become time-dependent, i.e.
\begin{equation}
\tilde{\mathcal Z}_k(t) ~=~ \tilde Z_k(t) \otimes \openone \,-\, \openone \otimes \tilde Z_k(t) ~.
\end{equation}
This facilitates the calculation of the average super-generator, and further allows the first-order effects of the RF fields upon the decoherence to be determined from the second-order terms in the cumulant expansion.
In contrast, assuming as usual that the random variables $\omega_k(t)$ have a mean value of zero at all times, it would be necessary to analyze the third-order terms in order to obtain these results.

Returning now to the problem of greatest interest here, in which there is only one noise generator which describes totally correlated decoherence as above and the corresponding random variable $\omega(t)$ is stationary and mean zero, the results given in Appendix A imply that the first two cumulants in the toggling frame are:
\begin{equation}
	\label{NMRcumulants2}	
\begin{array}{rcl}
	\tilde{\mathcal K}_1(t) & = & \displaystyle
	\frac{1}{t}\int_0^t dt'\, \langle {\mathcal L}_{det}(t')+\omega(t') \tilde{\mathcal Z(t')} \rangle ~=~ \frac{1}{t}\int_0^t dt'\, \tilde{\mathcal L}_{det}(t') \\[3ex]
	\tilde{\mathcal K}_2(t)  & = & \displaystyle
\frac{1}{t^2} \int_0^t dt_1 \int_0^{t_1} dt_2\, \Big( \big[ \tilde{\mathcal L}_{det}(t_1), \tilde{\mathcal L}_{det}(t_2) \big] ~+~2\, G(t_2-t_1)\, \tilde{\mathcal Z}(t_1)\, \tilde{\mathcal Z}(t_2)  \Big)
\end{array}
\end{equation}
In the last line we have introduced the autocorrelation function $G(\Delta t) =\langle \omega(t+\Delta t) \omega(t) \rangle$ for the stationary random noise variable $\omega(t)$.

\section{Refocusing noise with a Carr-Purcell sequence}
We now use these results to analyze an  implementation of a $\sigma_x^L$ rotation on a two-spin DFS qubit.
We will show that this implementation is applicable when the correlation time of the noise $\tau_c$ is long compared to the time required to apply a $\pi$-pulse to the spins .
It consists of a $(\pi/2)$-rotation of both spins in the DFS qubit about the $y$-axis, followed by a Carr-Purcell-style sequence consisting of an even number $2n$ of $\pi$-pulses separated by equal time intervals $\tau = t/2n$, and finally the inverse $(\pi/2)$-rotation, i.e.
\begin{equation}
\Big[\frac{\pi}{2}\Big]_y\Big( - \tau - \big[\pi\big]_x - \tau - \big[\pi\big]_x\Big)^{\!n}\, \Big[\frac{\pi}{2}\Big]_{\bar{y}}  \label{CPsequence}
\end{equation}
This transforms the weak $\sigma_z\sigma_z$ coupling between the two spins of the DFS qubit into $\sigma_x\sigma_x$, which projects to the $\sigma_x^L$ operator within the DFS (Eq. \eqref{EncodedOperations}).
Setting $\tau = \phi/(2n \pi J)$ thus yields a rotation by an angle $\phi$ around the logical $x$-axis.
Even though the state of the two spins is outside the DFS throughout the time $2n\tau$, the sequence of $\pi$-pulses is able to refocus the effects of the noise provided $\tau\ll\tau_c$.

Assuming instantaneous $\pi$-pulses, this follows from AHT since during any cycle $(0,2\tau)$ the internal Hamiltonian in the toggling frame $\tilde{H}_{int}$ alternates between $+\Delta\omega(\sigma_x^1-\sigma_x^2)+(\pi/2) J\sigma_x^1\sigma_x^2$ (in the interval $(0,\tau)$) and $-\Delta\omega(\sigma_x^1-\sigma_x^2)+(\pi/2) J\sigma_x^1\sigma_x^2$ (in the interval $(\tau, 2\tau)$), so that the zeroth-order average Hamiltonian is just $\bar{H}^{(0)}=(\pi/4)\sigma_x^1\sigma_x^2=(\pi/4)\sigma_x^L$.
This is in fact also the average Hamiltonian to all orders, since the toggling frame Hamiltonian commutes at all time, and the first cumulant is just the corresponding superoperator $\tilde{\mathcal K}_1 = \mathcal K_1 = (2\tau)^{-1} ({\bar H}^* \otimes \openone - \openone \otimes \bar H)$.

Again because the toggling frame Hamiltonians commute, the deterministic part of the Liouvillian $\tilde{\mathcal L}_{det}(t)$ does not contribute to $\tilde{\mathcal K}_2 = \mathcal K_2$ at the end of each cycle, nor at the end of the entire sequence.
The second cumulant is therefore determined by the stochastic part alone:
\begin{equation}
\label{2ndCum}
\mathcal K_2 ~=~ \frac2{(2n\tau)^2} \int_{0}^{2n\tau}\! dt_1\, \int_{0}^{t_1}\! dt_2\, G(t_2 - t_1) \tilde{\mathcal Z}(t_1) \tilde{\mathcal Z}(t_2) ~.
\end{equation}
Because each $\pi$-pulse simply changes the sign of $\tilde{\mathcal Z}(t)$ from the preceding interval, it follows that $\tilde{\mathcal Z}(t) = +\mathcal Z_x$ if $t$ is in an even interval $\big(2k\tau,\,(2k+1)\tau\big)$ and $\tilde{\mathcal Z} = -\mathcal Z_x$ if $t$ is in an odd interval $\big( (2k-1)\tau,\,2k\tau \big)$, where $k$ is an integer $0 \le k\le n$ and $\mathcal Z_x$ is the noise super-generator rotated along the $x$-axis. 
In addition, since the random variable $\omega(t)$ is stationary, the double integral over any two intervals $i$ and $i'$ will be equal to the double integral over any other pair $j$ and $j'$ providing that $|i - i'| = |j - j'|$.
These observations allow the overall double integral in Eq.~(\ref{2ndCum}) to be expressed as
\begin{equation}
\label{sumB}
\mathcal K_2 ~=~ \frac{2\,\mathcal Z^2_x}{(2n\tau)^2} \left( 2n A + \sum_{m=1}^{2n-1}\, (2n-m)\, B_m \right) ~\equiv~ \mathcal Z^2_x\, \zeta ~,
\end{equation}
where
\begin{equation}
\label{defA}
A ~\equiv~ \int_0^\tau\! dt_1 \int_0^{t_1}\! dt_2\, G(t_1 - t_2)
\end{equation}
and
\begin{equation}
\label{defB}
B_m ~\equiv~ {(-1)}^m\, \int_{m\tau}^{(m+1)\tau}\! dt_1 \int_0^\tau\! dt_2\, G(t_1 - t_2) 
\end{equation}
for $m = 1,\ldots,2n-1$.

In the case of Gaussian noise with autocorrelation function $G(t)=\Omega^2 e^{-t/\tau_c}$, one finds that
\begin{equation}\label{AGauss}
A ~=~ {(\Omega\tau_c)}^2\, \big( e^{-\tau/\tau_c} + \tau/\tau_c - 1 \big)
\end{equation}
and
\begin{equation}\label{Bbar}
B_m ~=~ \bar{B}{(-1)}^m\,  e^{-m\, \tau/\tau_c},\ \ \ \bar{B}=(\Omega\tau_c)^2e^{-\tau/\tau_c}(e^{\tau/\tau_c}-1)^2.
\end{equation}
On evaluating the geometric series in Eq.~(\ref{sumB}), one obtains the closed form
\begin{equation}
\label{zeta_in_closed_form}
\zeta ~=~ \frac{2\,\Omega^2\tau_c^2}{(2n\tau)^2} \bigg[ 2n \Big( \tau/\tau_c + e^{-\tau/\tau_c} - 1 \Big) +\bigg(\! \frac{1-e^{-\tau/\tau_c}}{1+e^{-\tau/\tau_c}}\bigg)^{\!\!2} \Big( 1 - 2n \big( 1 + e^{-\tau/\tau_c} \big) - e^{-2n\tau/\tau_c} \Big) \!\bigg] ,
\end{equation}
which is easily shown to go to zero as $\tau/\tau_c \rightarrow 0$  \footnote{In the limit $\tau/\tau_c\rightarrow \infty$ the behavior of $\zeta$ depends on the noise strength: If a constant noise strength is assumed,$ \zeta \rightarrow 0$ as $\frac{\Omega^2}{\tau/\tau_c}$. If instead we assume $\Omega\tau_c=$cst, $\zeta \rightarrow \infty$, since it is now $\zeta\propto \frac{\tau/\tau_c}{2n\tau^2}$, and the fidelity will go to zero.}.
We can quantify the protection afforded by the CP-sequence by taking the entanglement fidelity \cite{NielsenFid,DFSevan} of the superoperator with the ideal propagator for the sequence as a measure of its efficacy, $F\equiv\Tr{U_{\text{id}}^{-1} S}$.
Since the unitary part of the evolution commutes with the noise and gives the ideal propagator, the fidelity is just the trace of the superoperator, which for a single two-spin DFS qubit is
\begin{equation}
\label{fidelity}
F(\zeta) ~=~ \Tr{ e^{-\mathcal Z^2_x\zeta(2n\tau)^2/2} } ~=~ \tfrac1{8}\, \big( 3 + 4\, e^{-2\zeta 	n^2\tau^2}+ \, e^{-8\zeta n^2\tau^2}\big)  ~.
\end{equation}
The fidelity for cycles of CP-sequences of length $4$ and $16$ are plotted in Fig.~3.
As expected, it shows an improvement for an higher number of intervals and shorter time spacings with respect to the correlation time. 

\begin{figure}[htb]
	\begin{center}
		\includegraphics[scale=0.65]{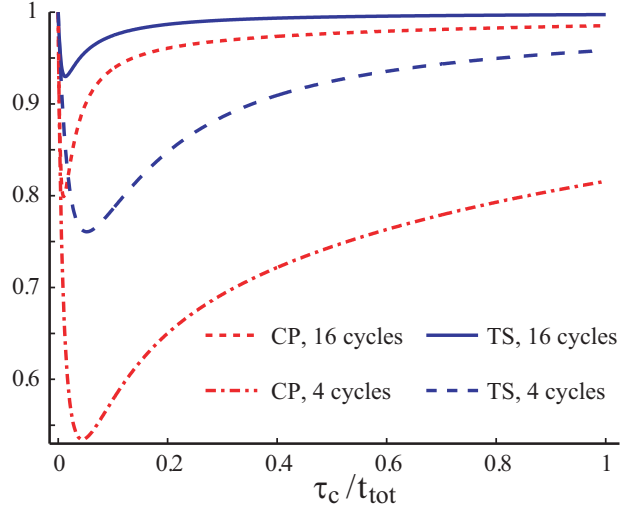}
	\end{center}
\caption{Gate fidelity as a function of the correlation time  for $4$ and $16$ cycles of the Carr-Purcell (CP)  and Time-Suspension sequences (TS).
The noise strength $\Omega$ was fixed at $1$ Hz., while the duration of the entire sequence was fixed at $t_{tot} = 4$ sec (where $t_{tot}=2n\tau$ for the CP sequence and $t_{tot}=4n\tau$ for the TS sequence).
The increase in fidelity at very short correlation times is due to the phase fluctuations becoming so fast that they produce essentially no effect at the given noise strength $\Omega$.}
	\label{fig:CPseqDecay}
\end{figure}

It is interesting to also consider a simple sequence that completely refocuses the internal Hamiltonian, namely
\begin{equation}
\Big( - \tau - \big[\pi\big]_x^1 - \tau - \big[\pi\big]_x^2- \tau - \big[\pi\big]_x^1 - \tau - \big[\pi\big]_x^2 \Big)^n, \label{TSsequence}
\end{equation}
where the superscripts on the pulse angles now refer to the spin affected by the pulses and $n$ is an integer.
This will be referred to in the following as the Time-Suspension (TS) sequence.
The average Hamiltonian is now zero, while the noise operator in the toggling frame is $\mathit{\tilde{Z}}_1=\pm(\sigma_z^1+\sigma_z^2)/2$ in the intervals 1 and 3 respectively and $\mathit{\tilde{Z}}_2=\pm(\sigma_z^1-\sigma_z^2)/2$ in the other two intervals.
If we sandwich the TS-sequence between a pair of $(\pi/2)$-pulses as we did for the CP, and again assume a stationary and Markovian Gaussian distribution of totally correlated noise, we find it is more effective at protecting the system from decoherence even when the number of $\pi$-pulses on each spin and the cycle time is the same, since the effective modulation rate is then faster (there is a pulse every $\tau/2$).
Indeed the relaxation superoperator for the TS-sequence is $\mathcal K_2=\zeta_1(\mathcal Z_1^2+\mathcal Z_2^2)+\zeta_2\, \mathcal Z_1\mathcal Z_2$  (see Appendix B), where $\mathcal Z_k = Z_k \otimes \openone - \openone \otimes Z_k$ ($k = 1,2$) and:
\begin{equation}\label{zeta1}
\begin{array}{ll}
\zeta _1=&\frac{\omega^2\tau_c^2}{16n^2\tau^2}
\left[
\left(\frac{1-e^{\tau/\tau_c}}{1+e^{2 \tau/\tau_c}} \right)^2
\left(e^{-4n\tau/\tau_c}\left(ne^{4\tau/\tau_c}-n+1\right)-1\right)e^{-3\tau/\tau_c}\right.\\
&\left.+2n\frac{\tau}{\tau_c}+ n\left(e^{-2\tau/\tau_c}-1\right)\left(2-e^{-\tau/\tau_c}\right) \right]
\end{array}
\end{equation}

\begin{equation}\label{zeta2}
\zeta_2=\frac{ \omega^2 \tau_c^2} {16 n^2\tau^2}
\frac{ \left(1-e^{\tau/\tau_c}\right)^2 e^{-4 \tau/\tau_c}}
{ 1+e^{2 \tau/\tau_c} } \left[e^{-4 n\tau/\tau_c}\left(ne^{4\tau/\tau_c}-n+1\right)+ne^{4 \tau/\tau_c}-(n+1)\right] 
\end{equation}
The fidelity is therefore:
\begin{equation}
\label{fidelity2}
\begin{array}{ll}
F(\zeta_1,\zeta_2) &~=~ \Tr{ \exp{(-\zeta_1(\mathcal Z_1^2+\mathcal Z_2^2)(4n\tau)^2/2-\zeta_2\mathcal Z_1\mathcal Z_2(4n\tau)^2/2)} } \\
&~=~ \frac{1}{2}\, e^{-\zeta_1(4n\tau)^2}\big(\cosh(-\zeta_1(4n\tau)^2)+ \cosh(-\zeta_2(4n\tau)^2/2) \big)  ~.
\end{array}\end{equation}

\section{Simulation of a selective DFS qubit gate}
The analytical expressions found above for the attenuation due to totally correlated noise with a stationary Gaussian Markov distribution apply only to the special case of ideal pulses (instantaneous in time), but similar behavior is expected under more realistic assumptions on the control fields. 
In particular, to act selectively only on some of the spins we would have to use the technique of SMP \cite{softpulses,PBEFFHMC:03}, thereby inducing a much more complex dynamics on the system for which closed form solutions are not available, but which can be studied via numerical simulations.

\begin{figure}[htb]
	\begin{center}
		\includegraphics[scale=1.0]{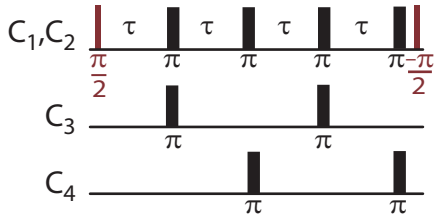}
	\caption{Selective rotation about the logical $x$-axis of a two-spin DFS qubit, while the evolution of a second DFS qubit under the internal Hamiltonian of the system is refocused.}
	\label{CPseq}
	\end{center}
\end{figure}

We have studied the accuracy with which a rotation about the logical $x$-axis can be performed by numerical simulations.
These simulations included the internal Hamiltonian, the external control Hamiltonian and totally correlated noise $\omega(t)$ with a stationary, Markovian Gaussian distribution.
The evolution was discretized into equal time steps, for each of which we calculated the propagator $U(t_k)=\exp(-i(\mathit{H}_{int}+\mathit{H}_{\textsf{rf}}(t_k)+\omega(t_k)\mathit{Z})\delta t)$.
The noise strength $\omega(t_k)$ is extracted from a multivariate gaussian probability distribution%
\endnote{This distribution was given by $\omega(t_k)=e^{-\delta t/\tau_c}\omega(t_{k-1})+r_k\sqrt{1-e^{-2\delta t/\tau_c}}$, where $r_k$ are normal distributed random numbers}, with a  covariance matrix  $C_{j,k}=\Omega^2e^{-|j-k|\delta t/\tau_c}$, where $j$ and $k$ are integers indicating the time intervals.
We then take the average of the superoperators $S_i=\bar{U_i}\otimes U_i$ obtained over a sequence of evolutions differing only by the random number seed.

We have performed one set of simulations using a fictitious two-spin molecule (chemical shift difference: $\Delta\omega=600$Hz, scalar coupling $J=50Hz$), and another using the internal Hamiltonian of $^{13}\mathsf C$-labeled crotonic acid, a molecule containing four carbon spins \cite{BFPTCH:02}.
Both sets of simulations were performed with instantaneous ideal pulses, and again with the strongly-modulating pulses used in actual NMR experiments. 
SMP are time-depend RF fields designed by a numerical search, and perform precise rotations of one or more spins while refocusing the evolution of all other spins in a molecule \cite{softpulses,PBEFFHMC:03}.

In the case of the two-spin molecule, since selective pulses are not required, we compare the results of SMP pulses with the dynamics under short,  collective pulses (called ``hard pulses'', $\pi$-pulse time $t_p=2\mu s$).
SMP appear to perform better even if they require longer times.
In the crotonic acid simulations, the sequence was designed not only to implement a selective $\pi/2$-rotation about the logical $x$-axis on the two spins in one DFS qubit, but to also refocus the evolution of the other two spins under the molecule's internal spin Hamiltonian (see Fig.~4).

The fidelities of these simulations are plotted as a function of correlation time in Fig.~5. 
Compared to simulations with ideal pulses, we observe a drop in the fidelity due to the finite duration of each pulse.
This drop is only in part accounted for by the increase in time in the cycle length.
Nevertheless, the effectiveness of the CP-sequence in preventing decoherence during the unavoidable excursions from the DFS is evident.  
\begin{figure}[htb]
	\begin{center}
		\includegraphics[scale=0.35]{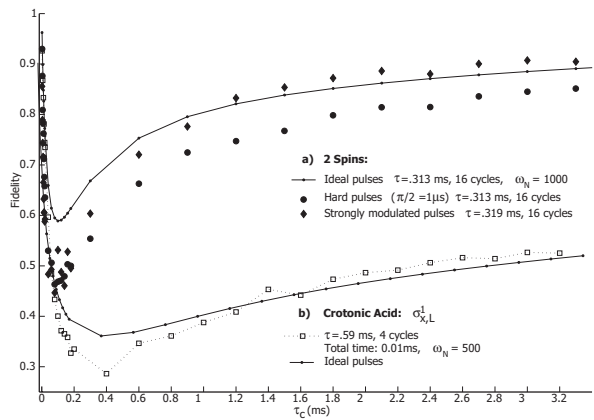}
	\caption{Fidelity for ideal and real pulses. a) Two-spin CP sequence implementing a $\pi/2$  rotation about the logical $\sigma_x$ (A fictitious spin system with $\Delta\omega=600$Hz and J=$50$Hz was used in the simulation) b) $\pi/2$ rotation about the logical $\sigma_x^1$ for Crotonic acid (see Ref. \cite{BFPTCH:02} for chemical shifts and J-coupling data). }
	\label{fig:CrotFid}
	\end{center}
\end{figure}

\section{Conclusions}
In this paper we have considered the difficulties of operating on quantum information stored in encoded qubits without losing the  protection from decoherence offered by the encoding.
Although we have focused upon the Hamiltonians and control fields operative in NMR for concreteness, similar difficulties will be encountered in other  functional realizations of quantum information processing, including squids, ion traps and quantum optics.
The most significant result is a demonstration that in many realizations, including NMR, the implementation of a universal set of quantum gates may be considerably simplified by briefly leaving the DFS while using dynamical decoupling to inhibit decoherence during these excursions.
This approach depends on the ability to operate on the system on time scales short compared to the correlation time of the noise.
In evaluating various possible realizations of quantum information processing,
it is important to characterize not only the decoherence rate, but also the spectral density of the underlying noise, to verify that the gate speed is sufficient to allow the noise to be refocused.
In addition to its role as a facile testbed for quantum information processing,
NMR spectroscopy provides widely applicable tools such as Average Hamiltonian theory by which one can calculate the efficacy of control sequences for refocusing the noise, and devise new ones for specific noise generators.

\bigskip\paragraph*{\hspace{-1.5em}Acknowledgments:} This work has been supported by the US Army Research Office under grant number DAAD19-01-1-0519, and by the Quantum Technologies Group of the Cambridge-MIT Institute.
	
\section{Appendix: The cumulant expansion}
We calculate the ensemble average of a time-ordered exponential in terms of the cumulant expansion.
First, expand the time-ordered average exponential $S=\langle \mathcal T \exp(-i \int_0^t dt' H(t'))\rangle$ via the Dyson series \cite{Sakurai:94}:
\begin{widetext}
\begin{equation}
\label{moments}
\begin{array}{lll}
S & = & \textbf{1}-i\int_0^t dt' \langle H(t')\rangle +
	\frac{(-i)^2}{2!} \mathcal T\int_0^t dt_1 \int_0^t dt_2 \langle H(t_1)H(t_2) \rangle + \cdots \\[2ex]
	&& \qquad\qquad +\, \frac{(-i)^n}{n!} \mathcal T \int_0^t dt_1 \cdots \int_0^t dt_n \langle H(t_1) \cdots H(t_n) \rangle +\cdots
 \end{array}
\end{equation}
\end{widetext}
The term $\langle H(t_1) \cdots H(t_n) \rangle$ is called the $n$-th moment of the distribution.
We want now to express this same propagator in terms of the cumulant function $K(t)$, defined by: 
\begin{equation}
\label{cumulant1}
S ~=~ e^{K(t)}
\end{equation}
The cumulant function itself can most generally be expressed as a power series in time:
\begin{equation}
\label{cumulant2}
K(t)=\sum_{n=1}^\infty \frac{(-i t)^n}{n!} K_n = -i t K_1+\frac{(-i t)^2}{2!}K_2 + \cdots
\end{equation}
Expanding now the exponential (\ref{cumulant1}) using the expression in equation (\ref{cumulant2}) we have:
\begin{equation}
\label{cumulant3}
\begin{array}{lcl}
S & = & \textbf{1}+K(t)+\frac{1}{2!}\left(K(t)\right) ^2 + \cdots \\[1ex]
& = &\textbf{1} -i t K_1 + \frac{(-i t)^2}{2!} (K_2+K_1^2) + \cdots
	\end{array}
\end{equation}
where in the second line we have separated terms of the same order in time. By equating terms of the same order in Eq. (\ref{cumulant3}) and (\ref{moments}) we obtain the cumulants $K_n$ in terms of the moments of order at most $n$.
For example:
\begin{equation}
\label{cum_mom}
\begin{array}{rl}
K_1 = &\displaystyle \frac{1}{t} \int_0^{t} dt'\, \big\langle H(t') \big\rangle \\[3ex]
K_2 = & \displaystyle \frac{1}{t^2}\; \mathcal T \int_0^t dt_1\int_0^t dt_2\, \big\langle H(t_1)H(t_2) \big\rangle - K_1^2
\end{array}
\end{equation}
The propagator can therefore be expressed in terms of the cumulant averages:
\begin{equation}
\label{cum_mom2}
\begin{array}{ccl}
      \big\langle H(t')\big\rangle_c	& = &\big\langle H(t')\big\rangle \\[1ex]
      \big\langle H(t_1)H(t_2) \big\rangle_c &	 = & \mathcal T\big\langle H(t_1)H(t_2) \big\rangle - \big\langle H(t_1)\big\rangle \big\langle H(t_2)
           \big\rangle
\end{array}
\end{equation}
The propagator can therefore be written as:
\begin{equation}
\label{cumulant5}
S ~=~ \exp\!\bigg(\! -i \int_0^t dt' \big\langle H(t') \big\rangle_c \,-\,\int_0^t dt_1\int_0^t dt_2\, \big\langle H(t_1)H(t_2) \big\rangle_c
    +\cdots\bigg)
\end{equation}

Note that if $H$ is a deterministic function of time, the ensemble averages can be dropped and $\big\langle H(t) \big\rangle_c = \int_0^t dt'\, H(t')$ becomes the time-average Hamiltonian, which is the first term in the Magnus expansion.
The second term in the cumulant expansion, on the other hand, becomes
\begin{equation}
\begin{array}{rcl}
&& \displaystyle \mathcal T \int_0^t dt_1\int_0^{t_1} dt_2\, H(t_1)H(t_2) - \bigg( \int_0^t dt'\, H(t') \bigg)^2 \\[3ex]
&=& \displaystyle 2 \int_0^t dt_1\int_0^{t_1} dt_2\, H(t_1)H(t_2) - \int_0^t dt_1  \int_0^t dt_2 \, H(t_1)H(t_2) \\[3ex]
&=& \displaystyle \int_0^t dt_1\int_0^{t_1} dt_2\, H(t_1)H(t_2) - \int_0^t dt_1  \int_{t_1}^t dt_2 \, H(t_1)H(t_2) \\[3ex]
&=& \displaystyle \int_0^t dt_1\int_0^{t_1} dt_2\, \big[ H(t_1),\, H(t_2) \big] ,
\end{array}
\end{equation}
where $[\cdot,\cdot]$ denotes the commutator and we have used the fact that the time-ordering operator $\mathcal T$ symmetrizes its argument with respect to permutation of the time points.
This is the second term in the Magnus expansion for the ``average'' (effective) Hamiltonian, and proceeding in this fashion one can in principle derive average Hamiltonian theory from the Dyson and cumulant expansions.

\section{Appendix: Cumulants of the Time Suspension sequence}	
We derive equations \ref{zeta1} and \ref{zeta2} for the fidelity attenuation of a Time Suspension sequence.
Consider first the basic cycle of the sequence, composed of four time intervals.
The average internal Hamiltonian is now zero, while the noise operator in the toggling frame is $\pm\mathcal{Z}_1=\pm(\sigma_z^1+\sigma_z^2)$ in the intervals 1 and 3 respectively and $\pm\mathcal{Z}_2=\pm(\sigma_z^1-\sigma_z^2)$ in the other two intervals.

\begin{figure}[htb]
	\begin{center}
		\includegraphics[scale=0.5]{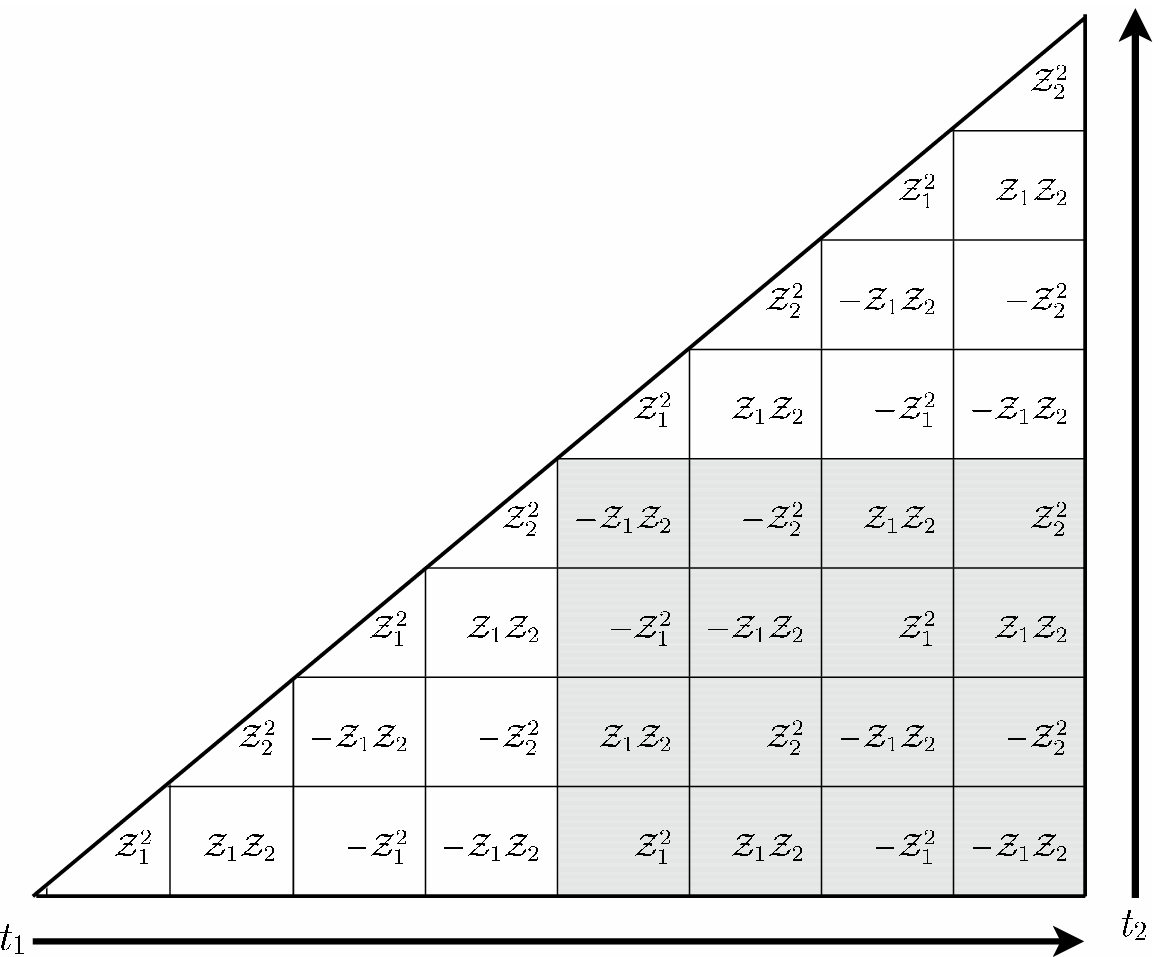}
	\caption{Domain of integration and toggling frame noise operator for the TS sequence}
	\label{integralplot}
	\end{center}
\end{figure}
By inspection of the domain of integration, the second order cumulant for the first cycle, (corresponding to the first triangle in figure \ref{integralplot}) is given by:
\begin{equation}
\begin{array}{ll}
\mathcal{K}_2^{(t)}&=\frac{\mathcal{Z}_1^2}{(4\tau)^2}(2A-\mathcal{Z}_1^2B_{3,1})+\frac{\mathcal{Z}_2^2}{(4\tau)^2}(2A-B_{4,2})+\frac{\mathcal{Z}_1\mathcal{Z}_2}{(4\tau)^2}(B_{2,1}-B_{4,1}-B_{3,2}+B_{4,3})
\\
&=\frac{\mathcal{Z}_1^2+\mathcal{Z}_2^2}{(4\tau)^2}(2A-\bar{B} e^{-2\frac{\tau}{\tau_c}})+
\frac{\mathcal{Z}_1\mathcal{Z}_2}{(4\tau)^2}\bar{B}(e^{-\frac{\tau}{\tau_c}}-e^{-3\frac{\tau}{\tau_c}})
\end{array}
\end{equation}
where $A$ and $\bar{B}$ where defined in equations \eqref{AGauss} and \eqref{Bbar} and  $B_{h,k}\equiv\bar{B}e^{-(k-h)\tau/\tau_c}$.
If the sequence is repeated $n$-times, we can divide the domain of integration in 4 by 4 time step squares and triangles (where the triangles are equivalent to the first cycle and the first of the possible squares is the shaded area in figure \ref{integralplot}).
Each triangle will give the same contribution calculated above for the first cycle(if there are $n$ cycles, we will have $n$ of them). The second cumulant from the first 4 by 4 square, corresponding to $t_1$  in the second cycle  and $t_2$ in the first cycle,  is:
\begin{equation}
\mathcal{K}_2^{(s)}=\frac{\mathcal{Z}_1^2+\mathcal{Z}_2^2}{(4\tau)^2}\bar{B}(2e^{-4\frac{\tau}{\tau_c}}- e^{-2\frac{\tau}{\tau_c}}- e^{-6\frac{\tau}{\tau_c}})+
\frac{\mathcal{Z}_1\mathcal{Z}_2}{(4\tau)^2}\bar{B}(e^{-3\frac{\tau}{\tau_c}}-e^{-\frac{\tau}{\tau_c}}+e^{-5\frac{\tau}{\tau_c}}-e^{-7\frac{\tau}{\tau_c}})
\end{equation}
To calculate the contributions from $t_1$ in cycle $k$   and $t_2$ in cycle $h$, it is enough to multiply this cumulant $\mathcal{K}_2^{(s)}$ by $e^{-4\frac{\tau}{\tau_c}(k-h)}$.
In general, for n cycles we obtain:
\begin{equation}
\mathcal{K}_2=\frac{1}{n^2}(n\mathcal{K}_2^{(t)}+\sum_{k=2}^n\sum_{h=1}^{k-1}e^{-4\frac{\tau}{\tau_c}(k-h)}\mathcal{K}_2^{(s)})\equiv \zeta_1(Z_1^2+Z_2^2)+\zeta_2Z_1Z_2
 \end{equation}
 with:
\begin{equation}
\begin{array}{ll}
\zeta _1=&\frac{\omega^2\tau_c^2}{16n^2\tau^2}
\left[
\left(\frac{1-e^{\tau/\tau_c}}{1+e^{2 \tau/\tau_c}} \right)^2
\left(e^{-4n\tau/\tau_c}\left(ne^{4\tau/\tau_c}-n+1\right)-1\right)e^{-3\tau/\tau_c}\right.\\
&\left.+2n\frac{\tau}{\tau_c}+ n\left(e^{-2\tau/\tau_c}-1\right)\left(2-e^{-\tau/\tau_c}\right) \right]
\end{array}
\end{equation}
and
\begin{equation}
\zeta_2=\frac{ \omega^2 \tau_c^2} {16 n^2\tau^2}
\frac{ \left(1-e^{\tau/\tau_c}\right)^2 e^{-4 \tau/\tau_c}}
{ 1+e^{2 \tau/\tau_c} } \left[e^{-4 n\tau/\tau_c}\left(ne^{4\tau/\tau_c}-n+1\right)+ne^{4 \tau/\tau_c}-(n+1)\right] 
\end{equation}

\bibliographystyle{apsrev}
\bibliography{version6}
\end{document}